\documentclass[aps,prl,reprint,numerical,superscriptaddress]{revtex4-1}

\usepackage[utf8]{inputenc}
\usepackage{amssymb}
\usepackage{amsmath}
\usepackage{multirow}

\usepackage[pdftex]{graphicx}
\usepackage{epstopdf} 
\usepackage{xspace}
\usepackage{dsfont}
\usepackage{bm}
\usepackage[normalem]{ulem}
\usepackage{hyperref}
\hypersetup{
    colorlinks,%
    citecolor=blue,%
    filecolor=blue,%
    linkcolor=blue,%
    urlcolor=blue
}

\providecommand{\ket}[1]{\vert #1\rangle} 
\providecommand{\bra}[1]{\langle #1\vert} 
\providecommand{\mean}[1]{\langle #1 \rangle} 

\usepackage{color}

\begin{document}

\title{Gate control of the spin mobility through the modification of \\ the spin-orbit interaction in two-dimensional systems}

\author{M. Luengo-Kovac}
\affiliation {Department of Physics, University of Michigan, Ann Arbor, Michigan 48109, United States}
\author{F. C. D. Moraes}
\affiliation {Instituto de F\'{i}sica, Universidade de S\~{a}o Paulo, S\~{a}o Paulo, SP 05508-090, Brazil}
\author{G. J. Ferreira}
\affiliation {Instituto de F\'{i}sica, Universidade Federal de Uberl\^{a}ndia, Uberl\^{a}ndia, MG 38400-902, Brazil}
\author{A. S. L. Ribeiro}
\affiliation {Instituto de F\'{i}sica, Universidade de S\~{a}o Paulo, S\~{a}o Paulo, SP 05508-090, Brazil}
\author{G. M. Gusev}
\affiliation {Instituto de F\'{i}sica, Universidade de S\~{a}o Paulo, S\~{a}o Paulo, SP 05508-090, Brazil}
\author{A. K. Bakarov}
\affiliation{Institute of Semiconductor Physics and Novosibirsk State University, Novosibirsk 630090, Russia}
\author{V. Sih}
\affiliation {Department of Physics, University of Michigan, Ann Arbor, Michigan 48109, United States}
\author{F. G. G. Hernandez}
 \email[Corresponding author.\\ Electronic address: ]{felixggh@if.usp.br}
\affiliation {Instituto de F\'{i}sica, Universidade de S\~{a}o Paulo, S\~{a}o Paulo, SP 05508-090, Brazil}

\date{\today}

\begin{abstract}
Spin drag measurements were performed in a two-dimensional electron system set close to the crossed spin helix regime and coupled by strong intersubband scattering. In a sample with uncommon combination of long spin lifetime and high charge mobility, the drift transport allows us to determine the spin-orbit field and the spin mobility anisotropies. We used a random walk model to describe the system dynamics and found excellent agreement for the Rashba and Dresselhaus couplings. The proposed two-subband system displays a large tuning lever arm for the Rashba constant with gate voltage, which provides a new path towards a spin transistor. Furthermore, the data shows large spin mobility controlled by the spin-orbit constants setting the field along the direction perpendicular to the drift velocity.  This work directly reveals the resistance experienced in the transport of a spin-polarized packet as a function of the strength of anisotropic spin-orbit fields.
\end{abstract}

\maketitle

The pursuit for a new active electronic component based on flow of spin, rather than that of charge, strongly motivates research in semiconductor spintronics \cite{wolf,zutic,awschalom,wunderlich,egues0}. Since the Datta-Das proposal for a ballistic spin transistor, full electrical control of the spin state was suggested using the gate-tunable Rashba spin-orbit interaction (SOI) \cite{rashba,datta,nitta,lechner,chuang}. Further studies, including the Dresselhaus SOI \cite{dresselhaus}, were made to assure a nonballistic transistor robust against spin-independent scattering \cite{schliemann1,ohno,kunihashi0}. For example, it has been demonstrated that SU(2) spin rotation symmetry, preserving the spin polarization, can be obtained in the persistent spin helix (PSH) formed when the strengths of the Rashba and Dresselhaus SOI are equal ($\alpha=\beta$) \cite{bernevig,koralek,walser,schliemann,khoda}. This is possible because the uniaxial alignment of the spin-orbit field suppresses the relaxation mechanism when the spins precess about this field while experiencing momentum scattering \cite{dyakonov}. Gate control of this symmetry point was experimentally observed \cite{khoda2,ishihara,eguesarxiv} and allowed to produce a transition to the PSH$^-$ ($\alpha=-\beta$) in the same subband \cite{ipsh}. Drift in those systems showed surprising properties \cite{yang0,kunihashi} such as the current-control of the temporal spin-precession frequency \cite{altmann}.  Although the helical spin-density texture could be even transported without dissipation under certain conditions \cite{bernevig}, the spin transport suffers additional resistance from the spin Coulomb drag \cite{amico1,amico2,tse,weber,yang}. These frictional forces appear as a lower mobility for spins than for charge and studies in new systems are still necessary to understand this important constraint for future devices.

A two-dimensional electron gas (2DEG) hosted in a quantum well (QW) with two occupied subbands offers unexplored opportunities for the study of spin transport \cite{fggh1,fggh2}. Theoretically, the inter- and intra-subband spin-orbit couplings (SOCs) have been extensively studied \cite{egues,egues1,egues2,egues3}. In terms of a random walk model (RWM) \cite{yangwalk}, the spin drift and diffusion was recently developed for these systems displaying two possible scenarios regarding the intersubband scattering (ISS) rate \cite{ferreira}. The interplay between the two subbands may introduce new features to the PSH dynamics, for example, a crossed persistent spin helix \cite{eguesprl} may arise when the subbands are set to orthogonal PSHs (i.e., $\alpha_1=\beta_1$ and $\alpha_2=-\beta_2$) in the weak ISS limit. In this report, we experimentally study spin drag in a system with the two-subbands individually set close to the PSH$^+$ and PSH$^-$, but with strong ISS, where the dynamics is given by the averaged SOCs of both subbands. The combination of long spin lifetime and high charge mobility allows us to determine the spin mobility and the spin-orbit field anisotropies with the application of an accelerating in-plane voltage. We are able to control the SOCs in both subbands and to show a linear dependence for the sum of the Rashba constants with gate voltage. Finally, we determine an inverse relation for the spin mobility dependence on the SOCs directly revealing the resistance experienced in the transport of a spin-polarized packet as a function of the strength of anisotropic spin-orbit fields.

\begin{figure}[ht!]
 \centering
 \includegraphics[width=1\columnwidth,keepaspectratio=true]{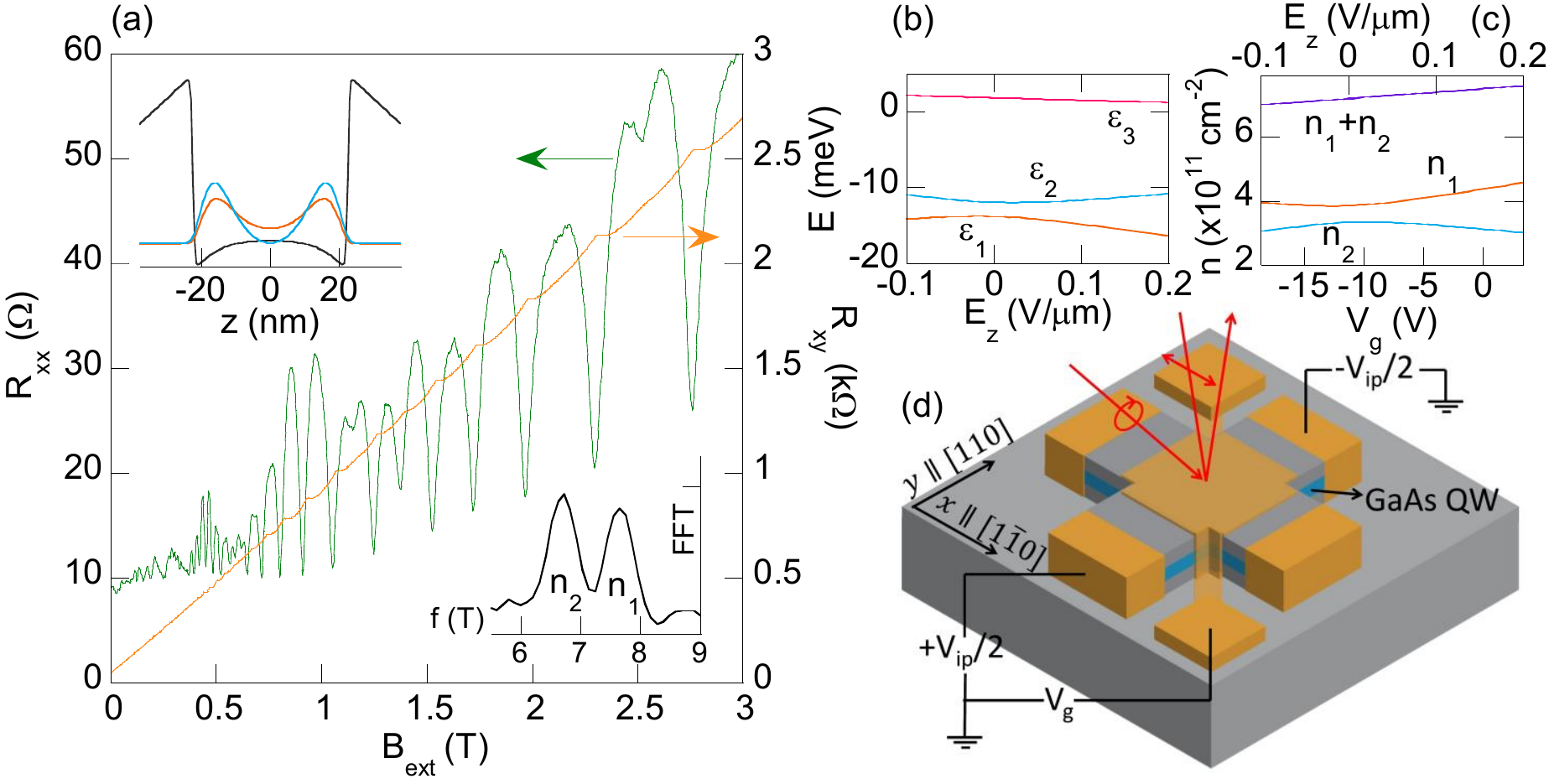}
 \caption{
 (a) Longitudinal ($R_{xx}$) and Hall ($R_{xy}$) magnetoresistance of the two-subband QW. From the SdH periodicity, one can obtain the subbands density $n_\nu$ in the lower inset. The top inset shows the potential profile and subbands charge density calculated from the self-consistent solution of Schr\"{o}dinger and Poisson equations for $E_z$=0. (b) Subband energy levels and (c) electron concentration dependence on V$_g$ and $E_z$. (d) Geometry of the device and contacts configuration.}
 \label{fig:1}
\end{figure}

The sample consists of a single 45 nm wide GaAs QW grown in the [001] ($z$) direction and symmetrically doped. Due to the Coulomb repulsion of the electrons, the charge distribution experiences a soft barrier inside the well. Figure \ref{fig:1}(a) shows the calculated QW band profile and charge density for both subbands. The electronic system has a configuration with symmetric and antisymmetric wave functions for the two lowest subbands with subband separation of $\Delta_{SAS}$ = 2 meV. The subband density (n$_1$= 3.7, n$_2$= 3.3$\times$10$^{11}$ cm$^{-2}$) was obtained from the Shubnikov-de Hass  (SdH) oscillations as shown in Fig. \ref{fig:1}(a) and the low-temperature charge mobility was 2.2$\times$10$^{6}$ cm$^{2}$/Vs \cite{bykov}. A device was fabricated in a cross-shaped configuration with width of $w$=270 $\mu$m and channels along the [1$\bar{1}$0] (x) and [110] (y) directions. Lateral Ohmic contacts deposited $l$=500 $\mu$m apart were used to apply an in-plane voltage ($V_{ip}$) in order to induce drift transport. For the fine tuning of the subband SOCs, a semitransparent contact on top of the mesa structure ($V_{g}$) was used to modify structural symmetry and subband occupation. The effect of $V_{g}$ on the subband energy levels ($\epsilon_\nu)$ and densities ($n_\nu$) is shown in Fig. \ref{fig:1}(b) and (c) as a function of the out-of-plane electric field ($E_z$). Note that the total density changes linearly with $V_{g}$ and that $V_{g}$=0 corresponds to a built-in electric field of 0.15 V/$\mu$m. Figure \ref{fig:1}(d) displays the experimental scheme with the connection of $V_{ip}$ and $V_{g}$ \cite{vip}. 

\begin{figure}[ht!]
 \centering
 \includegraphics[width=1\columnwidth,keepaspectratio=true]{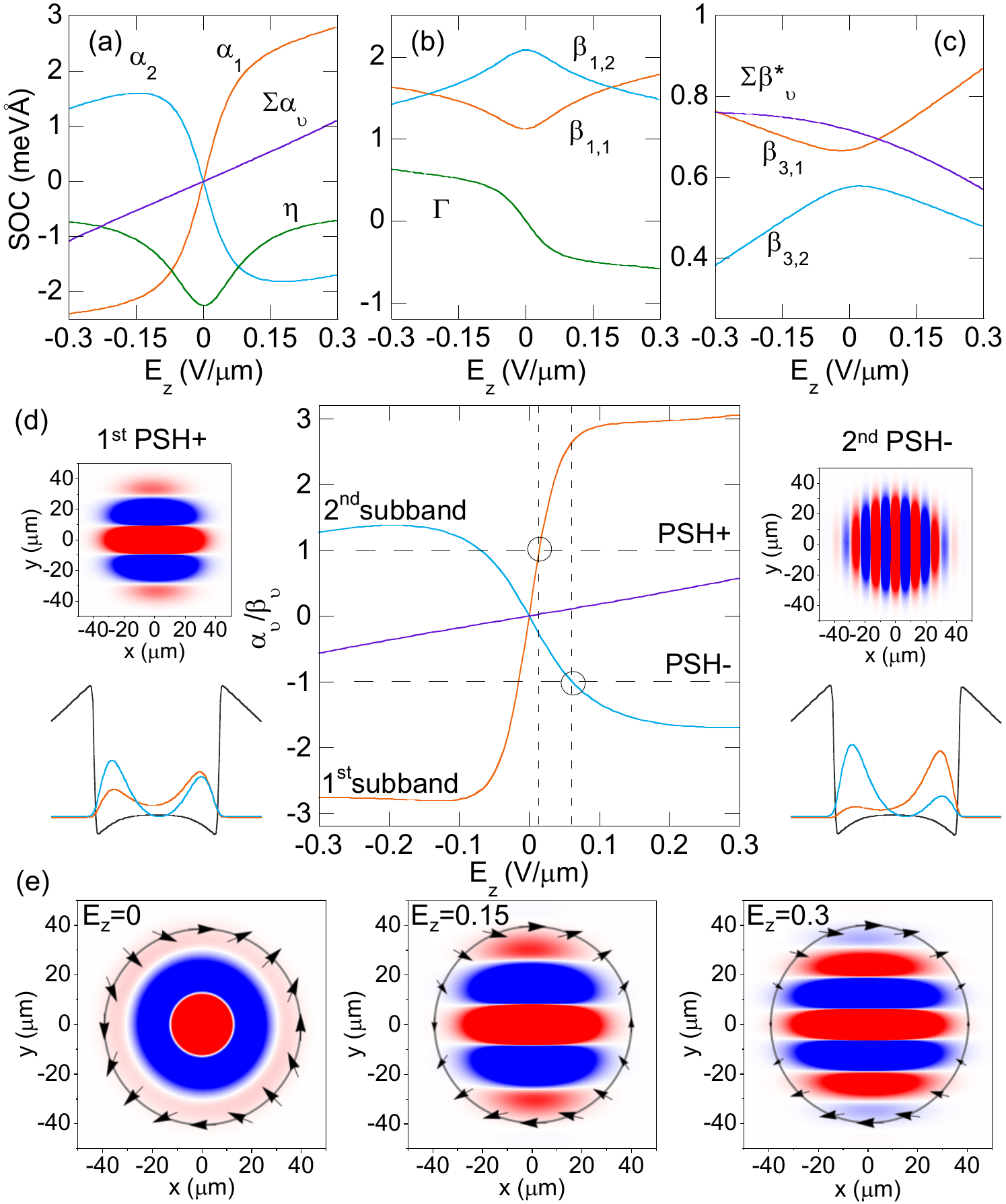}
 \caption{
 (a)-(c) Calculated SOCs for the Rashba ($\alpha_\nu$), linear ($\beta_{1,\nu}$) and cubic ($\beta_{3,\nu}$) Dresselhaus for each subband $\nu=\{1,2\}$, as well as intersubband SOCs $\eta$ and $\Gamma$ as a function of $E_z$.
 The purple lines give the sum of $\alpha_\nu$ and $\beta_\nu^*$.
 (d) The ratio $\alpha_\nu/\beta_\nu = \pm 1$ when the subband $\nu$ is set to the PSH$^\pm$ regime. 
 The insets show the single-subband magnetization maps on the $xy$ plane for the PSH$^\pm$ regimes, and the self-consistent potentials and subband densities for the respective $E_z$. (e) Two-subband magnetization maps in the strong ISS regime for different $E_z$. At $E_z = 0$ the well is symmetric ($\alpha_\nu = 0$) and the magnetization shows an isotropic Bessel pattern. For finite $E_z$ the broken symmetry leads to the stripped PSH pattern in accordance with the positive ratio $\sum \alpha_\nu / \sum \beta_{\nu}$ [purple line in panel (d)]. The arrows in the Fermi circle show the first harmonic component of $\sum \text{\bf{B}}_{SO,\nu}(\bm{k})$, illustrating the transition from isotropic to uniaxial field with increasing $E_z$. All the xy maps are frames of the spin pattern at t = 13 ns.
 }
 \label{fig:2}
\end{figure}

To describe the magnetization dynamics and the measured SO fields for our two-subband system, we combine the calculated SOCs with RWM \cite{yangwalk,ferreira,mamani}. For a [001] GaAs 2DEG, the x and y components of the SO fields for each subband $\nu = \{1,2\}$ are
\begin{equation}
 \text{\textbf{B}}_{SO,\nu}(\bm{k}) = \dfrac{2}{g\mu_B}
 \begin{pmatrix}
    \big(+\alpha_\nu + \beta_{1,\nu} + 2\beta_{3,\nu} \dfrac{k_x^2-k_y^2}{k^2}\big)k_y\\
    \big(-\alpha_\nu + \beta_{1,\nu} - 2\beta_{3,\nu} \dfrac{k_x^2-k_y^2}{k^2}\big)k_x
 \end{pmatrix},
\end{equation}
plus corrections due to the intersubband SOCs \cite{eguesprl,egues1,egues,egues2,eguesarxiv,egues3}. Above, $g=-0.44$ is the electron g-factor for GaAs and $\mu_B$ is the Bohr magneton. The SOCs are the usual Rashba $\alpha_\nu$ and linear $\beta_{1,\nu}$ and cubic $\beta_{3,\nu}$ Dresselhaus terms. Considering the strong intersubband scattering (ISS) regime of the RWM \cite{ferreira}, the randomization of the momenta $\bm{k}$ (within the Fermi circle $k=k_F$) and subband $\nu$ is much faster than the spin precession. Consequently, the dynamics is governed by an averaged SOC field $\mean{\bm{\text{\textbf{B}}}_{SO}} = (\mean{\text{B}_{SO}^x}, \mean{\text{B}_{SO}^y})$ transverse to the drift velocity $\text{\textbf{v}}_{dr} = (\text{v}_{dr}^x, \text{v}_{dr}^y)$ Namely, the field components read
\begin{align}
 \label{eqb1}
 \mean{\text{B}_{SO}^x} &= \Big[\dfrac{m}{\hbar g\mu_B} \sum_{\nu=1}^2 (+\alpha_\nu + \beta_\nu^*) \Big]\text{v}_{dr}^y,\\
 \label{eqb2}
 \mean{\text{B}_{SO}^y} &= \Big[\dfrac{m}{\hbar g\mu_B} \sum_{\nu=1}^2 (-\alpha_\nu + \beta_\nu^*) \Big]\text{v}_{dr}^x,
\end{align}
where $\beta_\nu^* = \beta_{1,\nu} - 2\beta_{3,\nu}$, and $m = 0.067m_0$ is the effective electron mass for GaAs and $\hbar$ is Planck's constant. Since $\text{B}_{SO}^{x(y)} \propto \text{v}_{dr}^{y(x)}$, it is convenient to analyze the linear coefficients $\text{b}^{x(y)} = \text{B}_{SO}^{y(x)}/\text{v}_{dr}^{x(y)}$, which are given by the terms between square brackets above.

The intra- and intersubband SOCs are calculated within the self-consistent Hartree approximation \cite{egues1,egues,egues2,egues3} for GaAs quantum wells tilted by $E_z$. The chemical potential is set to return the density $n = n_1+n_2 = 7\times 10^{11}$~cm$^{-2}$ for $E_z=0$, while it varies linearly for finite $E_z$ in Fig.~\ref{fig:1}(c). The SOCs are defined from the matrix elements $\eta_{\nu,\nu'} = \bra{\nu} \eta_w V'+\eta_H V_H' \ket{\nu'}$ and $\Gamma_{\nu,\nu'} = \gamma \bra{\nu}k_z^2\ket{\nu'}$, where $\ket{\nu}$ is the eigenket for subband $\nu$, $\eta_w = 3.47$~\AA$^2$ and $\eta_H=5.28$~\AA$^2$ are bulk coefficients \cite{egues1,egues,egues2,egues3,winkler}, $V' = \partial_z V(z)$ and $V_H' = \partial_z V_H(z)$ are the derivatives of the heterostructure and Hartree potentials along $z$, $\gamma = 11$~eV\AA$^3$ is the bulk Dresselhaus constant, and $k_z$ is the z-component of the momentum. The usual intrasubband Rashba and linear Dresselhaus SOCs are $\alpha_\nu = \eta_{\nu,\nu}$ and $\beta_{1,\nu} = \Gamma_{\nu,\nu}$. The non-diagonal terms are the intersubband SOCs $\eta = \eta_{12}$ and $\Gamma = \Gamma_{12}$. The calculated SOCs, plotted in Fig.~\ref{fig:2}(a)-(c) as a function of $E_z$, show agreement with previous studies \cite{studer,meier}. The high-density $n$ makes the cubic Dresselhaus $\beta_{3,\nu} \approx \gamma\pi n_\nu/2$ comparable with $\beta_{1,\nu}$, strongly affecting the PSH tuning \cite{walser} $\alpha_\nu = \beta_\nu$, with $\beta_\nu = \beta_{1,\nu}-\beta_{3,\nu}$.

Near $E_z \approx 0.04$~V/$\mu$m, the SOCs reach almost simultaneously the balanced condition for the PSH$^+$ in the first subband ($\alpha_1/\beta_1 = +1$) and for the PSH$^-$ in the second subband ($\alpha_2/\beta_2 = -1$), as shown by the ratio $\alpha_\nu/\beta_\nu$ in Fig. \ref{fig:2}(d). The expected magnetization patterns for the single-subband PSH is shown in the inset of Fig.~\ref{fig:2}(d). The PSH$^-$ shows more stripes than the PSH$^+$ due to the higher value of $\alpha$, which grows quickly within the $E_z$ range. However, the ratio of the averaged SOCs $(\sum \alpha_\nu)/(\sum \beta_\nu)$ approaches the PSH regimes only for $|E_z| > 0.3$~V/$\mu$m. As we will see next, the experimental data matches well the strong ISS regime of the RWM, therefore the dynamics is governed by the averaged SOCs. In this case, the expected magnetization patterns are shown in Fig.~\ref{fig:2}(e). With increasing $E_z$ the system transitions from isotropic ($E_z = 0$) to uniaxial ($E_z > 0.3$~V/$\mu$m), as indicated by the formation of stripes and the orientation of the first harmonic component of the total field $\sum \text{\bf{B}}_{SO,\nu}(\bm{k})$ [arrows in Fig.~\ref{fig:2}(e)].

\begin{figure}[ht!]
 \centering
 \includegraphics[width=1\columnwidth,keepaspectratio=true]{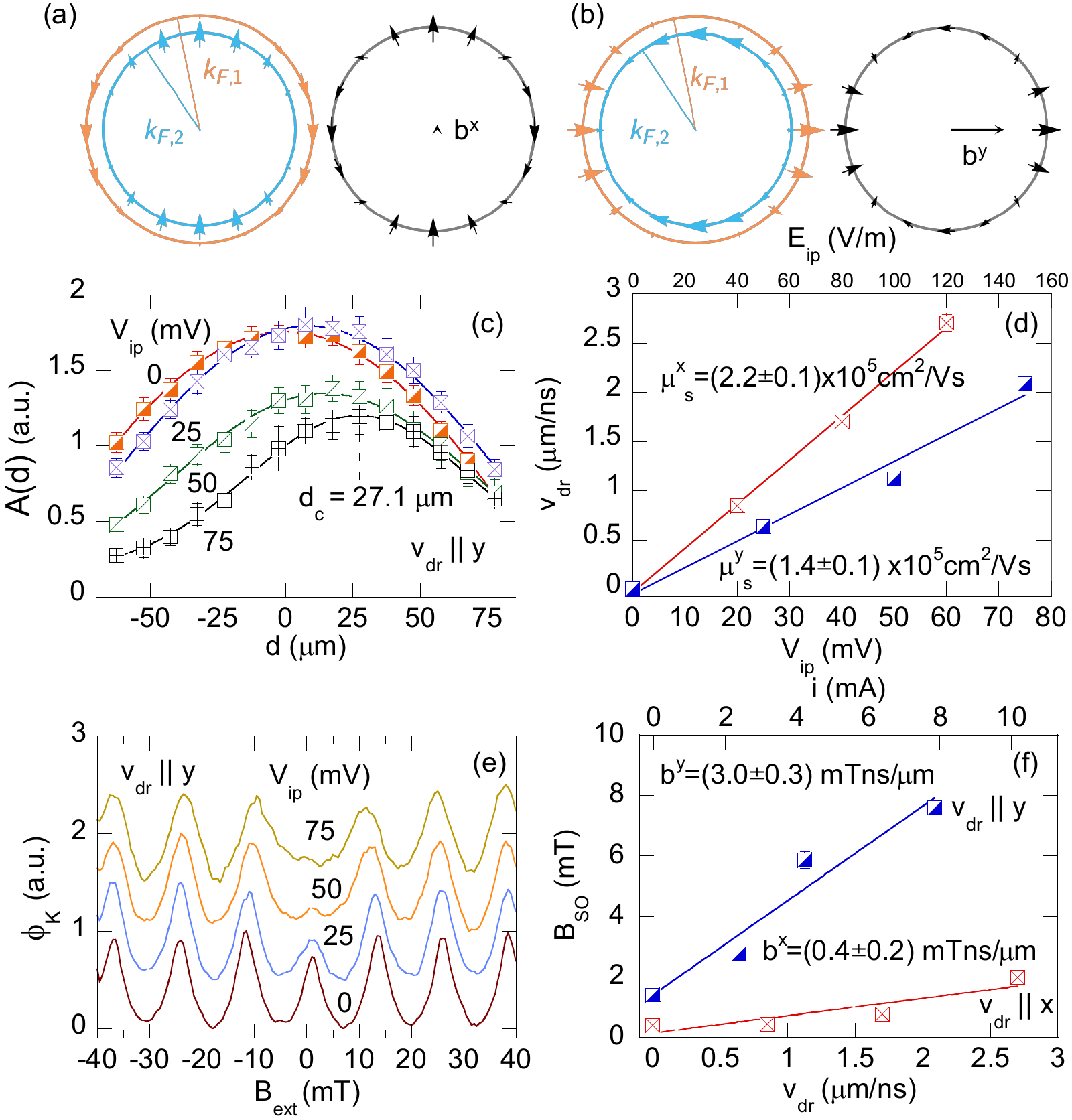}
 \caption{
Calculated coefficients $\bm{\text{b}}$ with {\bf v}$_{dr}$ parallel to (a) x and (b) y for each subband (colored) and total field (black). (c) Amplitude of the drifting spin polarization in space showing, for example, the center of the packet d$_c$ for 75 mV. (d) Linear dependence of v$_{dr}$ with the channel V$_{ip}$. The slope gives the spin mobility along v$_{dr}$ in x or y. (e) Field scan of $\phi_{K}$ for several V$_{ip}$ measured at d$_c$. (f) B$^{y(x)}_{SO}$ as function of v$^{x(y)}_{dr}$ and the current flowing in that channel. The slopes b$^{x(y)}$ give the strength of the SOCs that generate the field along y(x) for drift in x(y). The solid lines are gaussian (c) and linear (d and f) fittings. Scans taken at t=13 ns.
}
 \label{fig:3}
\end{figure}

We are interested in the determination of the anisotropy for the coefficients b$^{x(y)}$, estimated in one order of magnitude in Fig. \ref{fig:3}(a) and (b). We measured the spin polarization using time-resolved Kerr rotation as function of the space and time separation of pump and probe beams. All optical measurements were performed at 10 K. A mode-locked Ti:Sapphire laser with a repetition rate of 76 MHz tuned to 816.73 nm was split into pump and probe pulses. The polarization of the pump beam was controlled by a photoelastic modulator and the intensity of the probe beam was modulated by an optical chopper for cascaded lock-in detection. An electromagnet was used to apply an external magnetic field in the plane of the QW. The spatial positioning of the pump relative to the probe (d) was controlled using a scanning mirror. We defined the spin injection point to be x=y=0 at t=0. The application of an in-plane electric field (E$_{ip}$=V$_{ip}$/$l$), in the x or y-oriented channel, adds a drift velocity to the 2DEG electrons and allows us to determine the spin mobility and the spin-orbit field components \cite{kikkawa,kato,norman}.
 
The sample was rotated such that each channel under study was oriented parallel to the external magnetic field $\text{\bf{B}}_{ext}$$\parallel$$\text{\bf{v}}_{dr}$ for all measurements reported here. From the SOI form in k-space, we expected $\text{\bf{B}}_{SO}$$\perp$\text{\bf{v}}$_{dr}$ implying that the observable $\text{\bf{B}}_{SO}$ direction will be $\text{\bf{B}}_{SO}$$\perp$$\text{\bf{B}}_{ext}$. Considering this orientation, we can model the Kerr rotation signal as $\phi_{K}({\text{B}}_{ext},d)=\text{A}(\text{d})\cos\left(\omega\text{t}\right)$ with the precession frequency given by $\omega=(g\mu_B/\hbar)\sqrt{B_{ext}^2+B_{SO}^2}$, where A(d) is the amplitude at a given pump-probe spatial separation and B$_{SO}$ is the internal SO field component perpendicular to $\text{\bf{B}}_{ext}$ (and to $\text{\bf{v}}_{dr}$).

Figure \ref{fig:3} shows the results of the spin drag experiment with the gate contact open. Scanning the pump-probe separation in space at fixed long time delay (13 ns), we determined the central position d$_c$ of the spin packet amplitude for several V$_{ip}$ in a given crystal orientation. From the values of d$_c$ in Fig. \ref{fig:3}(c), we calculated the drift velocity as v$_{dr}$=d$_c$/t and plotted it as a function of V$_{ip}$ in Fig. \ref{fig:3}(d). The slope of the linear fit give us spin mobilities ($\mu^{x,y}_s$) in the range of 10$^5$ cm$^2$/Vs. Values in the same order of magnitude have been measured by Doppler velocimetry for the transport in single subband samples \cite{yang}. Nevertheless, in those systems the spin lifetimes were restricted to the picosecond range and the transport was limited to the nanometer scale. 

Following the drifting spin packet in space, Fig. \ref{fig:3}(e) displays a B$_{ext}$ scan from where changes in the amplitude of zeroth resonance determined B$_{SO}$ strength at d$_c$. As explained above, the data confirmed the perpendicular orientation between $\text{\bf{B}}_{SO}$ and $\text{\bf{v}}_{dr}$ and did not show a component parallel to $\text{\bf{B}}_{ext}$ within the experimental resolution \cite{kalevich}. From the Lorenztian shape of the B$_{ext}$ scan \cite{dzhioev,crooker}, we evaluated a spin lifetime of 7 ns at V$_{ip}$=0. This experiment was only possible due to the nanosecond spin lifetime in our sample that extends the spin transport to several tens of micrometers \cite{JAP,fggh3}.

Figure \ref{fig:3}(f) shows the fitted values of B$_{SO}$ for several V$_{ip}$ applied along x and y. We observed highly anisotropic spin-orbit fields in the range of several mT as expected from Fig \ref{fig:3}(a) and (b). The B$_{SO}$ orientation was aligned primary with the x axis in agreement with the simulation in Fig \ref{fig:2}(e). The slopes b$^{x(y)}$=B$^{y(x)}_{SO}$/v$^{x(y)}_{dr}$ give the strength of the SOCs that generate the field according to Eqs.~\ref{eqb1} and \ref{eqb2}. For this condition of the sample as-grown, we found $\sum\alpha_\nu$ = 0.57 $\text{meV\AA}$ and $\sum\beta^*_\nu$ = 0.75 $\text{meV\AA}$.

Note the inverse behaviour on V$_{ip}$ for the mobility and for B$_{SO}$ strength in perpendicular directions. In Fig. \ref{fig:3}(c) and (d), the axis with the largest mobility is also the axis with smaller spin-orbit field in the perpendicular direction. This result may be related to the spin Coulomb drag observed previously in the transport of spin-polarized electrons \cite{weber,yang}. Next, we demonstrate the direct control of the spin mobility through the gate modification of the subband SOCs. 

\begin{figure}[ht!]
 \centering
 \includegraphics[width=1\columnwidth,keepaspectratio=true]{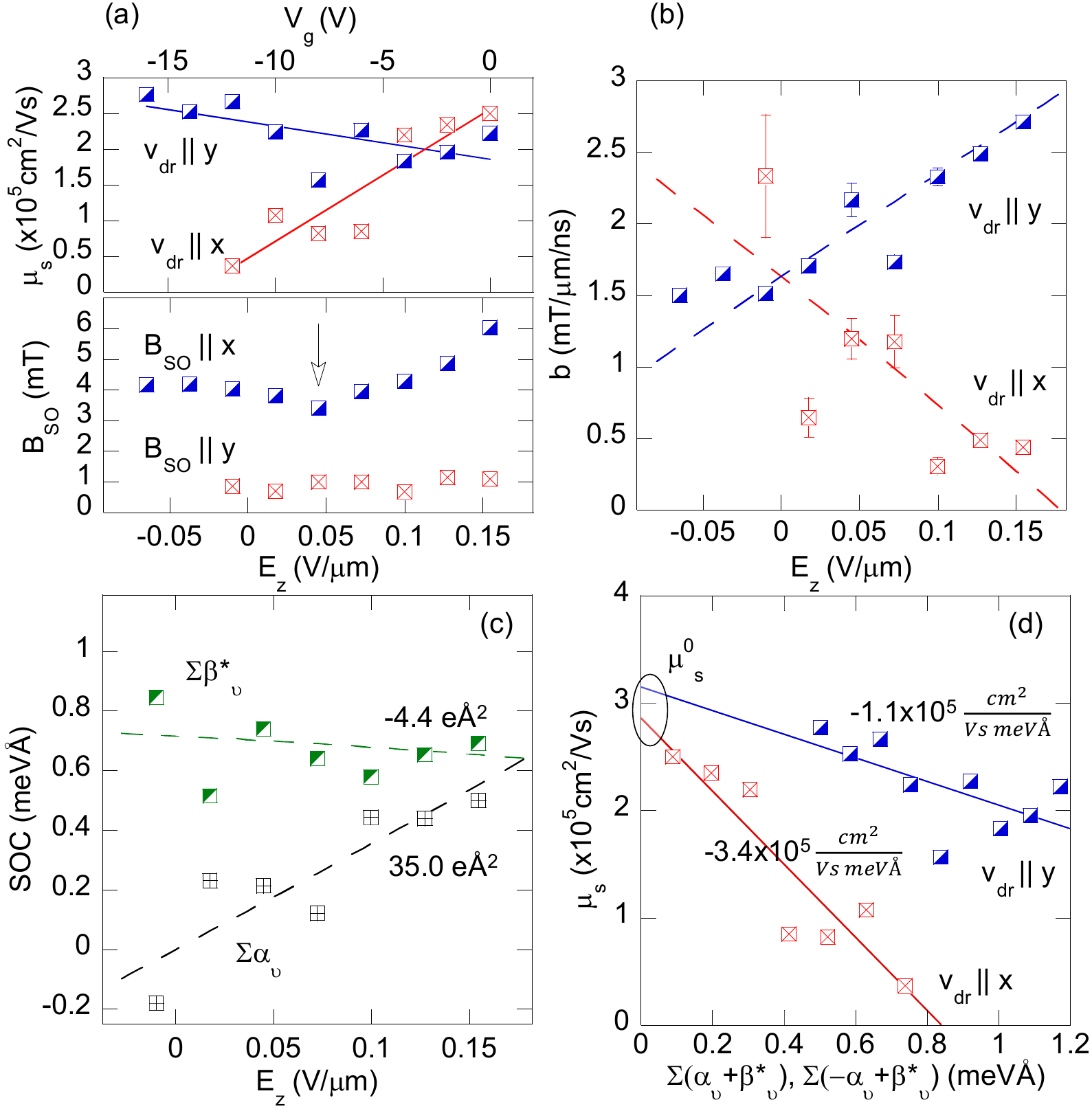}
 \caption{
(a) Spin mobility and B$_{SO}$ as a function of the gate-tunable E$_z$. (b) Ratio $b^{x(y)}$ from (a), showing a crossing at E$_z$=0. (c) SOCs obtained from the addition and subtraction of $b^x$ and $b^y$ in (b). (d) Spin mobility as function of the SOCs that define the B$_{SO}$ strength along the direction perpendicular to v$_{dr}$. The solid lines are linear fittings and the dashed lines (b,c) are the theoretical results from the RWM combined with the self-consistent calculation of the SOCs.}
 \label{fig:4}
\end{figure}

Figure \ref{fig:4}(a) shows that the magnitude and the orientation with the largest $\mu_s$ can be tuned by E$_z$. B$_{SO}$ displays anisotropic components with B$^{x}_{SO}$ being larger in all the studied range, which confirms the preferential alignment towards the PSH$^+$ in Fig. \ref{fig:2}(e). The variation of B$^x_{SO}$ has a minimum (indicated by an arrow) close to position when the second subband attains the PSH$^-$ (with $\text{\bf{B}}_{SO}$ along y). Dividing Fig. \ref{fig:4}(a) panels, the values for b are plotted in Fig. \ref{fig:4}(b). The lines plotted together with the data are the expected values using Eq. \ref{eqb1} and \ref{eqb2} with the SOCs from Fig. \ref{fig:2}(a)-(c). When the QW approaches the symmetric condition (E$_z$=0), b$^{x(y)}$ decreases removing the anisotropy of B$_{SO}$ as simulated in Fig. \ref{fig:2}(e). The addition and subtraction of $b^x$ and $b^y$ give the sum of the Rashba and Dressselhaus SOCs displayed in Fig. \ref{fig:4}(c). Dashed lines corresponding to the purple curves in Fig.\ref{fig:2}(a) and (c) are plotted together displaying excellent agreement. The slope for the Rashba SOI indicates a tuning lever arm of 35 e\AA$^2$. This value is considerably larger than those reported in recent studies for single subband samples, typically below 10 e\AA$^2$ \cite{eguesarxiv,walser}. Finally, Fig. \ref{fig:4}(d) presents $\mu^{x(y)}_s$ [from (a)] against the SOCs defining B$^{y(x)}_{SO}$: $\sum(-\alpha_\nu+\beta^*_\nu)$ and $\sum(\alpha_\nu+\beta^*_\nu)$, respectively. This last plot illustrates the inverse dependence, with negative slope, for the spin mobility and strength of the SOCs perpendicular to the drift direction. The different slopes for x and y channels give us a hint that this effect depends not only on how B$_{SO}$ changes with v$_{dr}$ (given by the SOCs) but also in the magnitude of the fields. A common maximum value $\mu^0_s$=3$\times10^5$cm$^{2}$/Vs was found independent of $\text{\bf v}_{dr}$ orientation.

In conclusion, we have studied a 2DEG system with two subbands set close to the crossed PSH regime under strong intersubband scattering and successfully described it using a random walk model. In the spin transport with nanosecond lifetimes over micrometer distances, we demonstrate the control of the subbands spin-orbit couplings with gate voltage and observed spin mobilities in the range of $10^5$cm$^{2}$/Vs. Specifically, the sum of the Rashba SOCs presents a linear behaviour with remarkably large tunability lever arm with gate voltage. We tailored the spin mobility by controlling the strength of the spin-orbit interaction in the direction perpendicular to the drift velocity. Our findings provided evidence of the rich physical phenomena behind multisubband systems and experimentally demonstrated relevant properties required for the implementation of a nonballistic spin transistor.
  
This work is a result of the collaboration initiative SPRINT No. 2016/50018-1 of the S\~{a}o Paulo Research Foundation (FAPESP) and the University of Michigan. F.G.G.H also acknowledges financial support from FAPESP Grants No. 2009/15007-5, No. 2013/03450-7, and No. 2014/25981-7 and 2015/16191-5. G.J.F. acknowledges the financial support from CNPq and FAPEMIG. The work at the University of Michigan is supported by the National Science Foundation under Grant No. DMR-1607779.


\begin{references}
\bibitem{vip} The measured low-temperature resistance was 10 $\Omega$ for the sample and 100 $\Omega$ for the lateral Ohmic contacts. Thus, the effective V$_{ip}$ in the channel is about 10 times smaller than the applied voltage. The values of V$_{ip}$ used for the plots and calculations are the effective ones.
\end{references}
\end{document}